\documentclass[twocolumn]{aa}
\usepackage{graphicx}
\usepackage{txfonts}
\begin{document}
\title{Probing the nature of Possible Open Cluster Remnants
with the Southern Proper Motion Program}
	
\author{Giovanni Carraro$^{1,2,3}$, Dana I. Dinescu$^2$, Terrence M. Girard$^2$, and William F. van Altena$^2$}

\offprints{Giovanni Carraro: gcarraro@das.uchile.cl}

\institute{
             $^1$Departamento de Astronom\'ia, Universidad de Chile, 
		 Casilla 36-D, Santiago, Chile\\
	     $^2$Astronomy Department, Yale University, P.O. Box 208101, 
		 New Haven, CT 06520-8101 USA\\
              $^3$Dipartimento di Astronomia, Universit\`a di Padova,
                 Vicolo Osservatorio 2, I-35122 Padova, Italy\\
             }

    \date{Received September 2004; accepted}

\abstract{We discuss the nature of eleven Possible Open Cluster Remnants (POCRs)
by using absolute proper motions from the Southern Proper Motion (SPM) 
Program 3 (Girard et al. 2004) combined with near infrared photometry from 2MASS.
The analysis is done by considering the distribution of stars 
in the Color-Magnitude and the Vector Point diagrams. We successfully probed 
the capabilities of the SPM catalog to detect a physical group by looking at the open
cluster Blanco~1. However,
within the uncertainties of the SPM3 catalog and  basing 
on 2MASS photometry we conclude that only one -ESO282SC26- out of eleven objects turns out to be a 
probable physical group. We suggest it is an open cluster 1.3 Gyr old and located 1.4 kpc
from the Sun. 
}

\authorrunning{Carraro et al.}
\titlerunning{Possible Open Cluster Remnant Candidates}
\maketitle
 %

\section{Introduction}
The recently published third release of the Southern Proper Motion (SPM3)
Program contains an almost complete catalog of absolute proper motions of stars
located in the southern hemisphere between declination of -20$^o$ and -45$^o$
down to V =17.5 (Girard et al. 2004)\footnote{http://www.astro.yale.edu/astrom/spm3cat/spm3.html}. 
The typical uncertainty in the proper motion
components is 4.0 mas yr$^{-1}$.
The proper motions are on the International Celestial Reference System by way of Hipparcos Catalog stars, 
and have an estimated systematic uncertainty of 0.4 mas yr$^{-1}$. \\

This catalog therefore represents a valuable tool to search for newly,
unknown open clusters like in the case of  the Hipparcos catalog (Platais et al. 1998) and
to investigate the 
nature of objects like the so-called Possible Open Cluster Remnants (POCRS),
proposed by Bica et al. (2001) to be candidate open clusters in advanced
stages of dynamical evolution, just before their dissolution
and merging with the general Galactic disk field.
These objects were selected through star counts, and are located rather 
high onto the Galactic plane (latitude larger than $15^o$),
where one does not expect to find any particular star over-density.
Bica et al. (2001) provide coordinates and diameters for a list of 34 clusters,
some of which have  NGC identification. \\

\noindent
Some of them have already been studied in details. NGC 6994 (M73) was proved
to be just a random enhancements of four bright stars by Carraro (2000) and
Odenchirken \& Soubiran (2002).
Recently, Villanova et al. (2003, 2004a) demonstrate that NGC 5385, NGC 2664 and Collinder 21 
are as well random alignments of field stars.
On the contrary, NGC 1901 was proved to be a genuine star cluster by Pavani et al. (2001)
and Villanova et al. (2004b).
In all cases the nature of the objects was clarified by carefully looking at
the distribution of proper motions and radial velocities, and therefore
at the kinematics of the candidate member stars.\\

\noindent
Since in most cases these objects are identified by clumps of bright stars,
we searched the SPM3 catalog and actually found proper motions for a significant 
number of stars (in the range 100-2000) in 11 POCRs extracted
from Bica et al. (2001) list (see Table~1 for details).
The SPM3 catalog together with absolute proper motions contains also
photographic B and V magnitudes, which unfortunately are affected by large errors, and 
therefore are not very useful to build up good Color Magnitude Diagrams (CMDs).
However, the SPM3 catalog has been combined with the 2MASS (Skrutskie et al. 1997) one, thus providing for
nearly all the objects near-infrared J, H and K photometry.\\
An analysis similar to the one presented here was done by Baumgardt (1998),  
who was able to clarify the nature of 8 controversial open clusters by analyzing
their parallaxes, proper motion components and B and V magnitudes from the Hipparcos catalog, and
by Baumgardt et al. (2000) and Dias et al. (2001, 2002a).\\

\noindent
The analysis of the clusters is performed by comparing the proper motion distribution
and the CMDs of the stars in the POCR area and in an off-set field (typically
half a degree apart). The main aim of this study is to clarify the nature
of these POCRs, and in case we found that one object is a probable physical cluster,
we would provide for it the first estimate of its fundamental parameters, and a list
of candidate members for spectroscopic follow-up.\\

 \noindent
The layout of this paper is as follows.\\
In Sect.~2 we test SPM3 capabilities against
the open cluster Blanco~1.
In Sect.~3 we describe the analysis method, and
in Sect.~4 we discuss individually all the candidates.
Our conclusions are finally highlighted in Sect.~5.

 \begin{table*}
 \tabcolsep 0.6cm
\caption{Basic parameters of the objects under investigation.
 Coordinates are for J2000.0 equinox and have been 
taken from Dias et al. (2002b)}
 \begin{tabular}{cccccc}
\hline
 \hline
 \multicolumn{1}{c}{Name} &
 \multicolumn{1}{c}{$RA$}  &
 \multicolumn{1}{c}{$DEC$}  &
 \multicolumn{1}{c}{$l$} &
 \multicolumn{1}{c}{$b$} & 
 \multicolumn{1}{c}{$Diam$}\\
 \hline
 & $hh:mm:ss$ & $^{o}$~:~$^{\prime}$~:~$^{\prime\prime}$ & [deg.] 
& [deg.] & arcmin\\
 \hline
ESO486SC45      & 05:16:43 & -24:02:17 & 226.07 & -30.80 &  3.5\\
NGC 1891        & 05:21:25 & -35:44:24 & 239.69 & -32.87 & 15.0\\
NGC 1963        & 05:32:17 & -36:23:30 & 240.99 & -30.86 & 13.0\\
ESO424SC25      & 05:49:49 & -32:28:20 & 237.72 & -26.37 &  9.0\\
ESO425SC06      & 06:04:50 & -29:10:59 & 235.39 & -22.28 &  6.0\\
ESO425SC15      & 06:14:35 & -29:22:30 & 236.37 & -20.35 &  6.0\\
ESO437SC61      & 10:48:03 & -29:23:26 & 273.06 &  26.22 &  5.0l\\
ESO442SC04      & 12:34:05 & -29:24:38 & 298.40 &  33.30 & 11.0\\
IC 1023         & 14:32:25 & -35:48:13 & 324.95 &  22.71 &  5.0\\
ESO282SC26      & 19:13:52 & -42:38:58 & 335.01 & -21.89 & 15.0\\
ESO464SC09      & 20:59:37 & -29:23:12 &  15.92 & -39.43 &  4.0\\
 \hline
Blanco 1        & 00:04:07 & -29:50:00 &  15.57 & -79.26 &  90\\
 \hline\hline
 \end{tabular}
 \end{table*}

\begin{figure}
 \centering
\includegraphics[width=9cm]{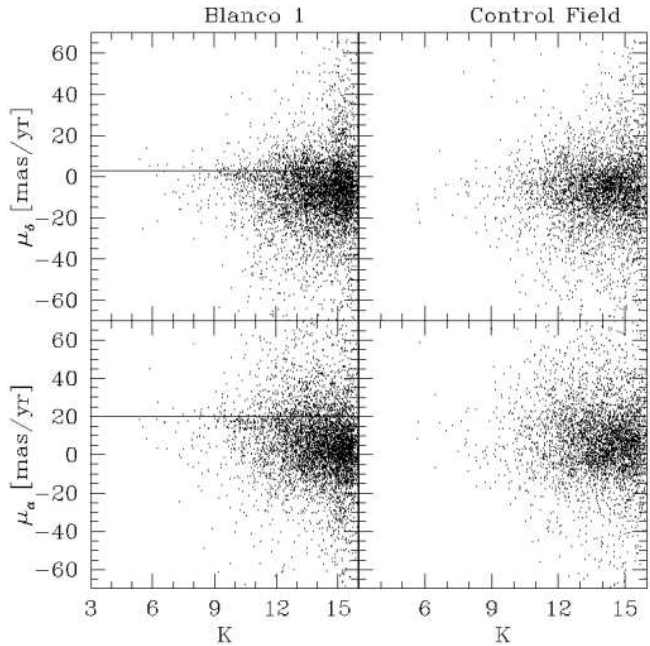}
\caption{Trend of proper motion components versus K mag. in Blanco~1
(left panels) and a nearby field (right panels). The solid line is
Blanco~1 mean proper motion components taken from Dias et al. (2001)}
\end{figure}

 \begin{figure}
 \centering
 \includegraphics[width=9cm]{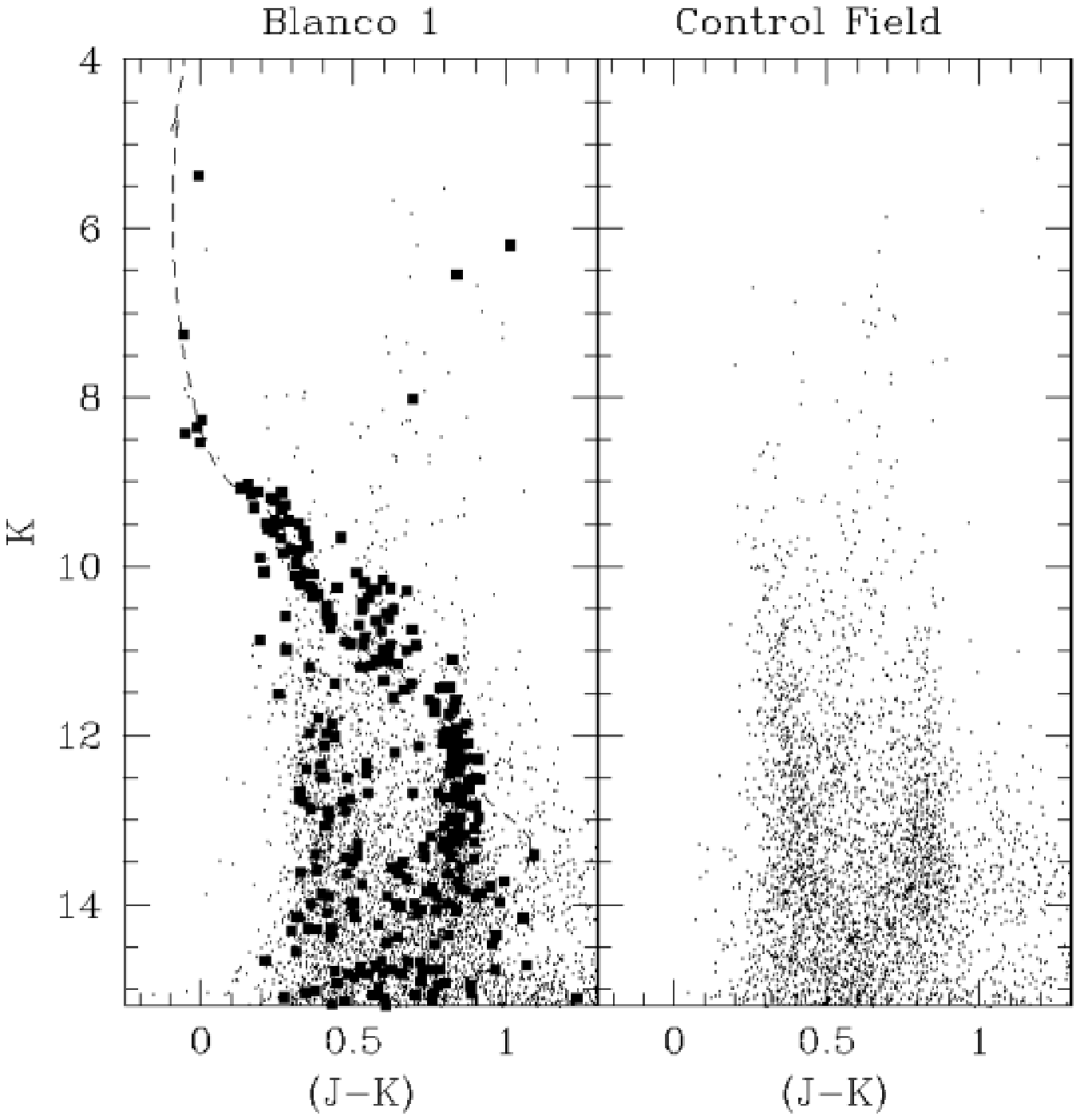}
 \caption{{\bf Left panel:} CMD of Blanco~1 from 2MASS. Probable members are
shown as filled circles. The dashed line is the Schmidt-Kaler (1982) ZAMS
shifted by the cluster distance modulus and reddening.
{\bf Right panel:} CMD of a nearby field from 2MASS.}
 \end{figure}

 \begin{figure}
 \centering
 \includegraphics[width=9cm]{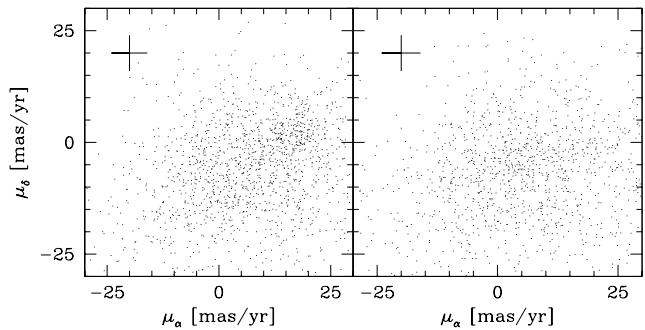}
 \caption{{\bf Left panel:}VPD from SPM3 for all the stars brighter than K= 13.5
in the field of Blanco~1. {\bf Right panel:} The same as left panel, but for
 the nearby field. The cross indicates the 1$\sigma$ error bar from SPM3.}
 \end{figure}

\section{Testing SPM3 capabilities: the open cluster Blanco~1}
All the fields here considered are located high onto the galactic plane,
and therefore several galaxies are easily encountered. We removed all the 
extended sources from the catalog before commencing the analysis.\\

\noindent
To probe the capabilities of SPM3 to detect a physical group, we searched
for a well known, genuine open cluster in the catalog, and we found
the open cluster Blanco 1. 
This is a young nearby cluster, 290 pc from the Sun,
with a diameter of 2.5 deg., and with mean proper motions
$\mu_\alpha~=~20.06\pm0.49$ and $\mu_\delta~=~3.44\pm0.25$ mas yr$^{-1}$ 
(Van Leeuwen 1999, Dias et al. 2001).\\

\noindent
We extracted from the SPM3 catalog 6225 stars in an area of 2.5 squared deg.
centered on Blanco~1 nominal center (see Table~1), and 4677 stars in a similar size area
5 degrees northward of the cluster.
The results are shown in Fig.~1, where we present the trend of proper motion components
versus K mag. The left panels show the distribution of Blanco 1 stars, while
the right panels show the distribution of stars in the off-set field.
The solid lines in the left panels are the Hipparcos mean proper motion
component (Van Leeuwen 1999).
First of all we note that the proper motion components start to be significantly
scattered at K $\approx$ 11.0 . 
However, we clearly see a significant stars concentration  at $9.0 \leq K \leq 12.0$
in both the proper motion components, close to the expected mean motion
component of the cluster from Hipparcos (solid lines), although a
sizeable difference is visible in the case of $\mu_\alpha$. 
These stars probably constitute the cluster Blanco 1. \\

\noindent
Since the precision (1$\sigma$) of the SPM3 catalog is about 4mas yr$^{-1}$,
we picked up from the cluster plots all the stars which lie within a 
strip 16 mas yr$^{-1}$ wide (2$\times \sigma$ ) centered at Blanco~1
proper motion mean values, where we expect to find 95\% of member stars
within our uncertainties.
The result is shown in Fig.~2, where we plot the 2MASS CMD of Blanco~1.
In the left panel we plot all the stars detected in the cluster area,
whereas in the right panel we plot all the stars found in the control field.
The stars selected on a proper motion basis are plotted in the left panel
as filled circles, and they clearly define a nice Main Sequence (MS) with
some scatter.
The dashed line in this plot is to guide the eye, and it is a 
Zero Age Main Sequence (ZAMS) from Schmidt-Kaler (1982) 
for the distance ($m-M = 7.1$) and the reddening (E(J-K) = 0.019) of
Blanco~1.\\

 \begin{figure}
 \centering
 \includegraphics[width=9cm]{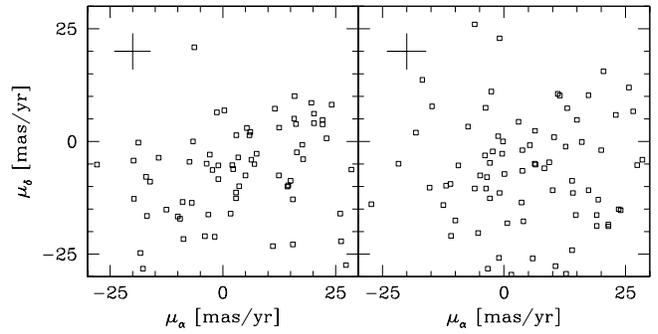}
 \caption{{\bf Left panel:}VPD from SPM3 for all the stars brighter than K= 13.5
in the smaller field of Blanco~1. {\bf Right panel:} The same as left panel, but for
 the smaller area nearby field. The cross indicates the 1$\sigma$ error bar from SPM3.}
 \end{figure}

\begin{figure}
 \centering
 \includegraphics[width=9cm]{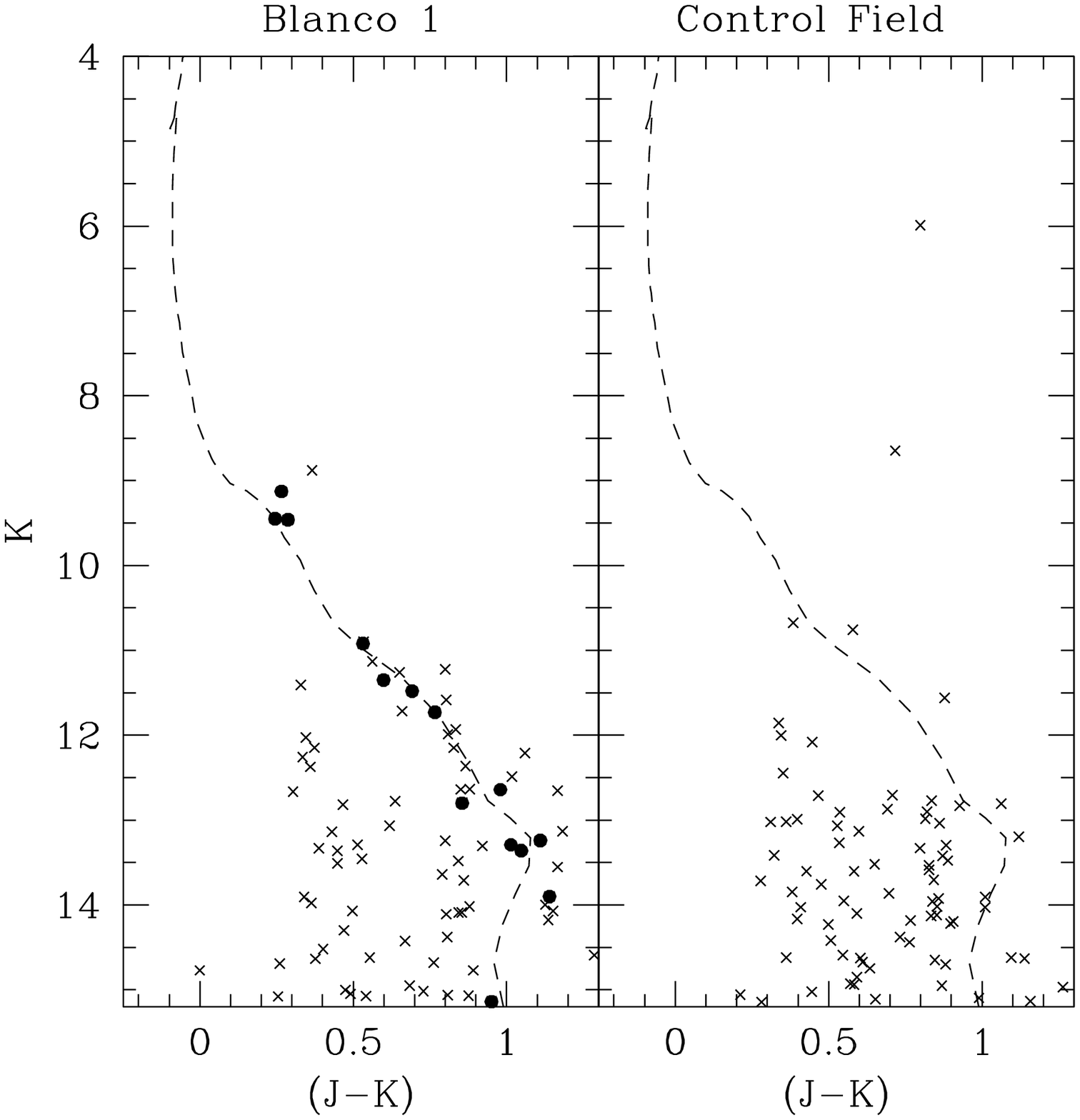}
 \caption{{\bf Left panel:} CMD of Blanco~1 from 2MASS in an area of
$15^{\prime} \times 15^{\prime}$.Probable members are
shown as filled circles. The dashed line is the Schmidt-Kaler (1982) ZAMS
shifted by the cluster distance modulus and reddening.
{\bf Right panel:} CMD of a nearby same area field from 2MASS.}
 \end{figure}

\noindent
Finally, in Fig.~3 we show the VPD for the stars brighter than K = 13.5
in the Blanco~1 field (left panel) and in the off-set field (right panel).
The cut off in magnitude is mainly motivated by the need to limit the field
star contamination; however it is the population brighter than this cut-off
that actually better defines the cluster in Fig.~2.
In the VPD the cluster readily appears as a concentration of stars
centered at  $\mu_\alpha \approx 20.$ and $\mu_\delta \approx 3.$,
and this clustering does not have a counterpart in the off-set field.
If we imagine to remove the cluster, the stars distribution in the VPD is the
same both for the cluster and for the off-set field.\\

\noindent
However, since in general OCRs are less populated ensembles of stars and occupy
smaller areas than Blanco~1, we decide to further check SPM~3 capabilities on
a smaller area in the open cluster Blanco~1, with the aim to mimick the case
of an OCR. We searched the SPM~3 catalogue in a $15^{\prime} \times 15^{\prime}$
area centered in Blanco~1, and we extracted 101 stars in the cluster area,
and 108 stars from a same area off-set field.\\
The results are shown in Figs.~4 and 5, where we plot the VPD and the CMD,
respectively. Even considering a smaller area and a significanly smaller number
of stars, the open cluster Blanco~1 is readily detected as a clump of stars
around the mean proper motions in the VPD (Fig.~4), and as a poorly sampled MS in the CMD
(Fig.~5).
We derived the mean proper motion for the probable member stars (14, in this smaller field), 
and they turned out
to be $\mu_\alpha =19.63\pm3.44$ (1 standard deviation) and $\mu_\delta =4.21 \pm 4.14$
(1 standard deviation).
These values are in good agreement with Van Leeuwen 1999 ones, although the 
dispersion is larger, but well consistent with the SPM~3 uncertainties.\\

\noindent
All these plots demonstrate that the absolute proper motions from SPM3 combined
with 2MASS photometry are capable to identify a physical group of stars
with the same tangential motions.\\
\noindent
Therefore it provides us with a tool to be used when looking for suggested
physical groups like POCRs.

\section{Analysis of POCRs: the method}
All the targets of the present study are over-densities of bright 
stars with respect to the surrounding
field (Bica et al 2001).
This is a necessary condition for a bound group of stars to exist.\\
However, only if these stars share common motion and the common motion stars exhibit distinctive
features in the CMD, we are going to conclude that the stars actually form a physical system
(Platais et al 1998).\\
On the contrary, if neither they  show common motion nor known features in the CMD, we must
conclude that these stars are simple random star accumulations along the line of sight.\\

\noindent
In the following we are going to apply the same technique as for Blanco 1 to the POCRs
listed in Table~1. 
In details, we compare the target field with a control field 
taken in all the cases at a distance of half a degree
from the nominal target center.\\

\noindent
At odds with Blanco~1,  here we do not know the object proper motions, and therefore
we look for possible member
within $\pm$8~mas yr$^{-1}$ from the mode of the proper motion components distribution, 
both in the target and in the control field.
This way, we hope to catch most of the object members, if any. 
The searching area depends on the cluster radius, as listed in 
Table~1, but in all cases is much larger than the cluster diameters as proposed by Bica et al. (2001).\\

\begin{figure}
\centering
\includegraphics[width=9cm]{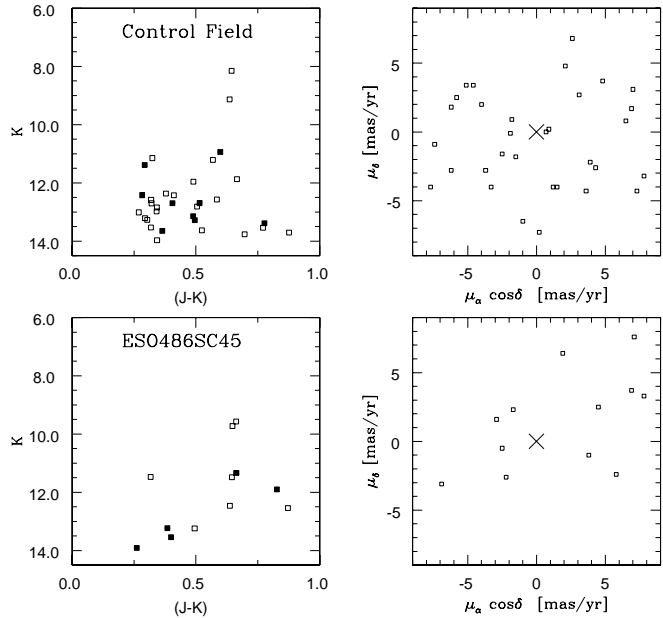}
\caption{The ESO486SC45 group. Only stars within 2 SPM3 $\sigma$
from the proper motion distributions mode are plotted. {\bf Bottom left panels:} 2MASS CMD.
Filled symbols indicate stars with proper motions within 1 $\sigma$
from the proper motion distributions mode, open symbols stars with proper motion within 2 $\sigma$
from the proper motion distributions mode.
{\bf Bottom right panel:} VPD; the cross is centered in the field proper motions mode.
{\bf Upper left panel:} Same as the bottom left panel, but for the control field;
{\bf Upper right panel:} Same as the bottom right panel, but for the control field.
} 

\end{figure}
\begin{figure}
\centering
\includegraphics[width=9cm]{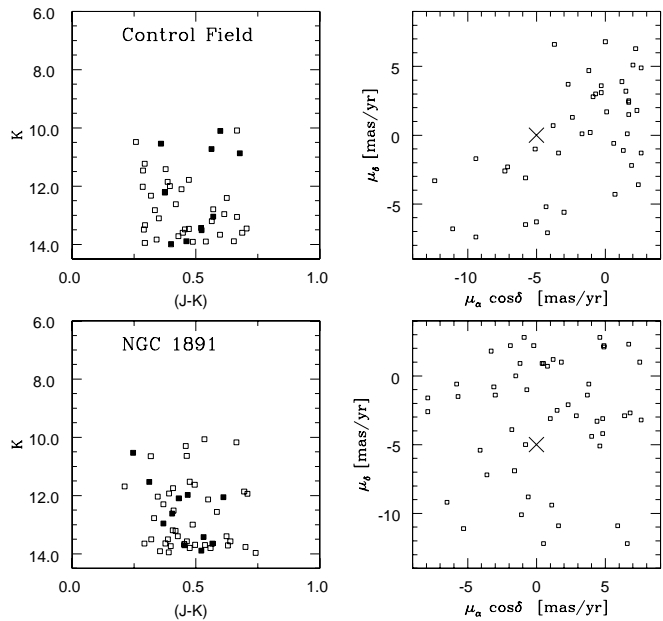}
\caption{Same of Fig.~6, but for NGC 1891} 
\end{figure}

\begin{figure}
\centering
\includegraphics[width=9cm]{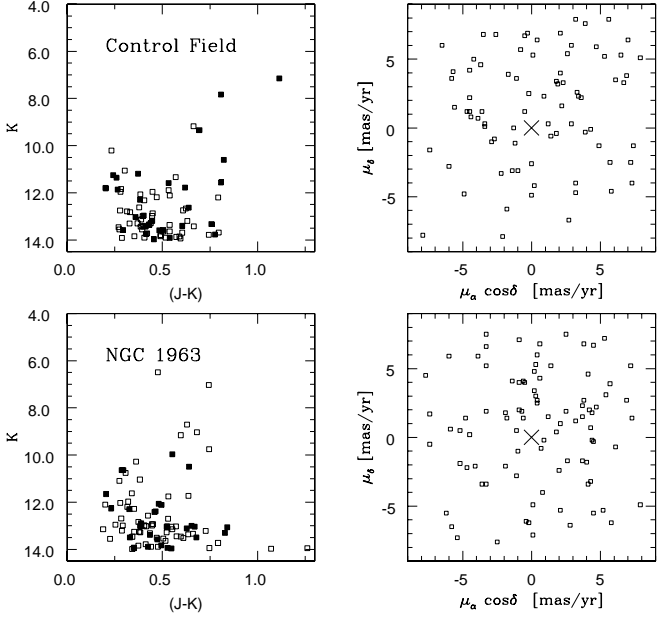}
\caption{Same of Fig.~6, but for NGC 1963} 
\end{figure}

\noindent
In the following we are showing how each POCR candidate and the corresponding
off-set field look like
in the VPD and CMD. This is to show the reader how
we basically decided upon the probable presence of a cluster. 
Basically we search for possible differences in the proper motions between target and field,
and look for distinctive features in the CMD of common motions stars.\\

We searched in the SPM3 catalogue for stars brighter  than $ K=14$, and considered only the stars
with proper motions within 2 SPM3 $\sigma$ (8~mas yr$^{-1}$) from the mode of the proper motion components 
distribution computed
taking into account all the detected stars.\\ 
The adopted limiting magnitude allow us to detect
the brightest candidate stars and limit the field star contamination, which increases at increasing magnitude.
Down to K=14 the 2MASS photometry is very reliable, and we do not expect to find many OCR members
at fainter magnitudes, since the samll mass stars quickly escape the cluster.
Besides, the magnitude range of OCR candidate stars is very variable, some of them having very bright stars,
and other having fainter stars which produce the overdensity above the Galactic field in a way that
one can hope also to find a few more faint members.

\section{Individual objects analysis}
In this Section we provide for each object
some details of the searching procedure and 
the corresponding VPDs and  CMDs for the target and control fields.

\subsection{ESO486SC45}
This object is shown in Fig.~6. We considered a $15^{\prime} \times 15^{\prime}$ 
area, and we found 142 stars in the object field, and 154 stars in the control field.
The mode of the proper motion distribution turned out to be around
$\mu_\alpha~=~0.0$ and $\mu_\delta~=~0.0$ mas yr$^{-1}$ both in the target
and in the control field, and the control field 
looks less scattered than the cluster.
By inspecting Fig.~6, we do not see any of the features which would lead us to 
think about the presence of a cluster, namely neither a  concentration in the VPDs
which look very scattered, nor a sequence in the CMDs. 
In this case the field seems to have more common motion stars
than the target.\\ 
We therefore conclude that ESO486SC45 is not a physical group.

\subsection{NGC~1891}
This target is presented in Fig.~7. We considered a $30^{\prime} \times 30^{\prime}$
area, and we found 379 stars in the target, and 386 stars in the offset field. 
The mode of the proper motion distribution  turned out to be about
$\mu_\alpha~=~0.0$ and $\mu_\delta~=~-5.0$ mas yr$^{-1}$ in the target,
 and 
$\mu_\alpha~=~-5.0$ and $\mu_\delta~=~0.0$ mas yr$^{-1}$ in the field.  
By inspecting Fig.~7, we see that the CMDs of target and field stars are
similar, with the same spread in color and without any distinctive features.
No relevant concentrations are detected in the VPDs.\\
We therefore propose that NGC 1891 is not a physical group.

\subsection{NGC~1963}
This target is presented in Fig.~8. We considered a $30^{\prime} \times 30^{\prime}$
area, and we found 523 stars in the target, and 437 stars in the offset field.
The mode of the  proper motion distribution out to be
$\mu_\alpha~=~0.0$ and $\mu_\delta~=~0.0$ mas yr$^{-1}$ both in the target and in the field.
By inspecting Fig.~8,  we readily see that distribution of the stars 
both the CMDs and the VPDs are very
similar, with basically the same scatter in proper motions and color.\\
Therefore  we suggest that NGC 1963 is not a physical group.

\begin{figure}
\centering
\includegraphics[width=9cm]{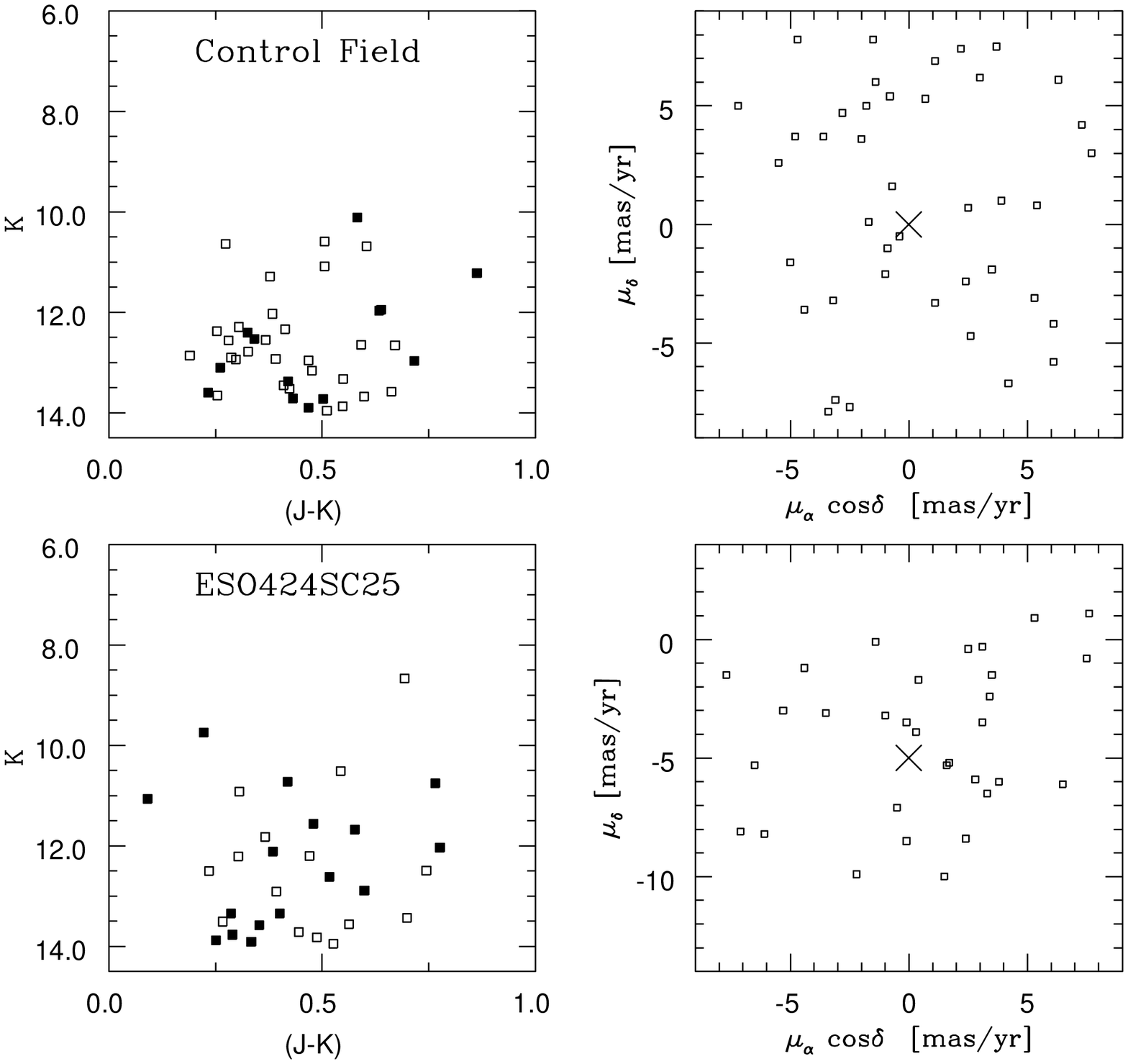}
\caption{Same of Fig.~6, but for ESO424SC25} 
\end{figure}

\subsection{ESO424SC25}
This object is shown in Fig.~9. We considered a $20^{\prime} \times 20^{\prime}$
area, and we found 172 stars in the target, and 244 in the control field.
The mode of the  proper motion distribution turned out to be
$\mu_{\alpha} \approx 0.$ mas yr$^{-1}$ and $\mu_{\delta} \approx -5.$ mas yr$^{-1}$ in the target and 
$\mu_\alpha~=~0.0$ and $\mu_\delta~=~0.0$ mas yr$^{-1}$ in the field.
From the analysis of Fig~9, we can conclude that ESO424SC25 is clearly 
not a physical group since both the CMDs and the VPDs are very scattered and
look similar.

\begin{figure}
\centering
\includegraphics[width=9cm]{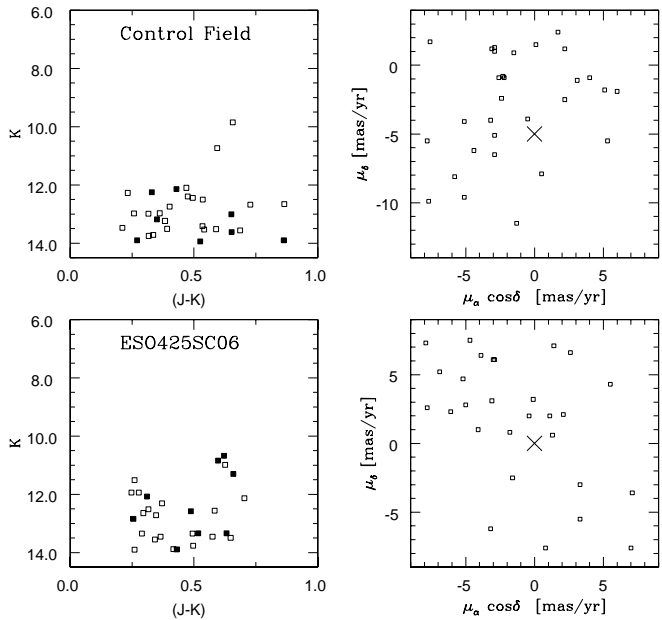}
\caption{Same of Fig.~6, but for ESO425SC06} 
\end{figure}

\begin{figure}
\centering
\includegraphics[width=9cm]{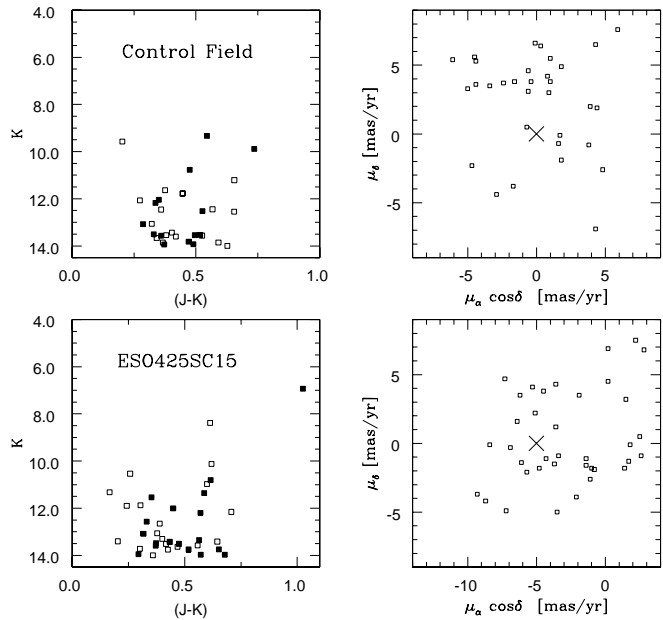}
\caption{Same of Fig.~6, but for ESO425SC15} 
\end{figure}

\begin{figure}
\centering
\includegraphics[width=9cm]{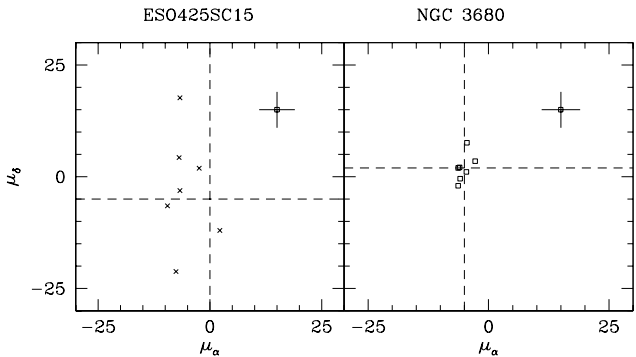}
\caption{VPD for ESO425SC15 and NGC 3680 from Tycho~2. The cross indicate the 2 $\sigma$
error bars. Dashed lines represent the mean proper motions of the two targets.} 
\end{figure}

\subsection{ESO425SC06}
This object is presented in Fig.~10. We considered a $15^{\prime} \times 15^{\prime}$ 
area, and we found 194 stars in the target area, and 198 stars in the off-set field.
The mode of the  proper motion distribution out to be
$\mu_{\alpha} \approx 0.0$ mas yr$^{-1}$ and $\mu_{\delta} \approx 0.0$ mas yr$^{-1}$ in the target and 
$\mu_\alpha~=~0.0$ and $\mu_\delta~=~-5.0$ mas yr$^{-1}$ in the field.
From the analysis of Fig~10, ESO425SC06 exhibits the same scatter as the control field
both in the CMD and in the VPD with the lack of any concentration or particular feature, 
and therefore we suggest it is  not a physical group.

\subsection{ESO425SC15}
This target is compared in Bica et al. (2001) with NGC~3680, a very
well known open cluster (Anthony-Twarog \& Twarog 2003) 
which exhibits a significant depletion
of low mass stars in the MS. Actually the two targets
look very similar on the plane of the sky (Figs.~2 and 3 in Bica et
al. 2001), thus leading to the possibility that ESO425SC15
might be a genuine cluster.
This target is depicted in Fig.~11. We considered a $15^{\prime} \times 15^{\prime}$ 
area, and we found 213 stars in the target area, and 214 stars in the nearby field.
The mode of the  proper motion distribution out to be
$\mu_{\alpha} \approx -5.0$ mas yr$^{-1}$ and $\mu_{\delta} \approx 0.0$ mas yr$^{-1}$ in the target and 
$\mu_\alpha~=~0.0$ and $\mu_\delta~=~0.0$ mas yr$^{-1}$ in the field.
From the analysis of CMDs and VPDs we conclude that there is not cluster
in the direction of ESO425SC15.
As a confirmation of this finding, we plot in Fig.~12 the VPDs from Tycho~2
( H{\o}g et al. 2000)
of the 7 brightest stars which defines the concentration and compare it with 
the same number of bright stars in NGC~3680. 
It is readily visible that 
ESO425SC15 is a chance alignment of different proper motion stars.

\subsection{ESO437SC61}
This target is depicted in Fig.~13. 
We considered a $15^{\prime} \times 15^{\prime}$ 
area, and found 133 stars in the target, and 160 in the control field.
The mode of the  proper motion distribution turned out to be
$\mu_{\alpha} \approx -5.0$ mas yr$^{-1}$ and $\mu_{\delta} \approx 0.0$ mas yr$^{-1}$ in the target and 
$\mu_\alpha~=~0.0$ and $\mu_\delta~=~-5.0$ mas yr$^{-1}$ in the field, and both distributions
are very scattered. As a result, and from the appearance of CMDs and VPDs we suggest
that this group is not a physical system.

\subsection{ESO442SC04}
This target is discussed  in Fig.~14. We considered a $20^{\prime} \times 20^{\prime}$ 
area, and we found 283 stars in the target, and 226 stars in the nearby field.
The mode of the  proper motion distribution turned out out to be
$\mu_{\alpha} \approx -5.0$ mas yr$^{-1}$ and $\mu_{\delta} \approx -5.0$ mas yr$^{-1}$ in the target and 
$\mu_\alpha~=~-5.0$ and $\mu_\delta~=~-10.0$ mas yr$^{-1}$ in the field.
Also in this case, we cannot see any particular feature in the VPDs and CMDs,
which do not look different, and therefore suggest that 
ESO442SC04 is not a physical system.

\subsection{IC~1023}
This target is shown in Fig.15. We considered a $15^{\prime} \times 15^{\prime}$ 
area, and we found 432 stars in the target area, and 415 stars in the off-set field,
roughly the same number of stars.
The mode of the  proper motion distribution turned out to be about
$\mu_{\alpha} \approx -5.0$ mas yr$^{-1}$ and $\mu_{\delta} \approx -5.0$ mas yr$^{-1}$ in the target and 
$\mu_\alpha~=~-5.0$ and $\mu_\delta~=~-5.0$ mas yr$^{-1}$ in the field.
By inspecting Fig.~15, we see that the stars are scattered both in the field and in the target.
No distinctive features are present in the CMDs, and we must conclude also in this case
that we are not facing any physical group.

\begin{figure}
\centering
\includegraphics[width=9cm]{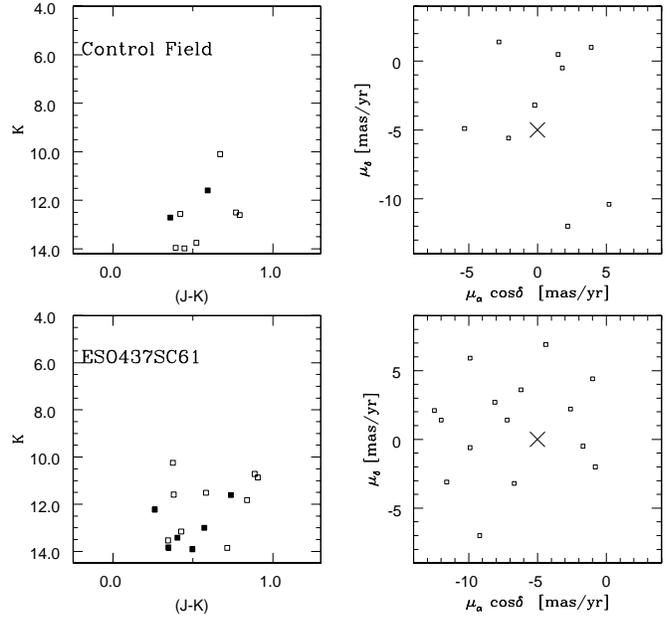}
\caption{Same of Fig.~6, but for ESO437SC61} 
\end{figure}

\begin{figure}
\centering
\includegraphics[width=9cm]{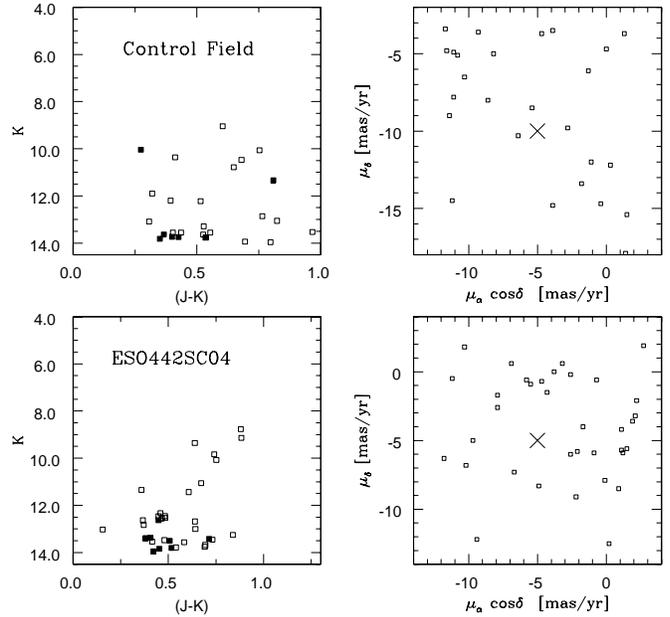}
\caption{Same of Fig.~6, but for ESO442SC04} 
\end{figure}

\begin{figure}
\centering
\includegraphics[width=9cm]{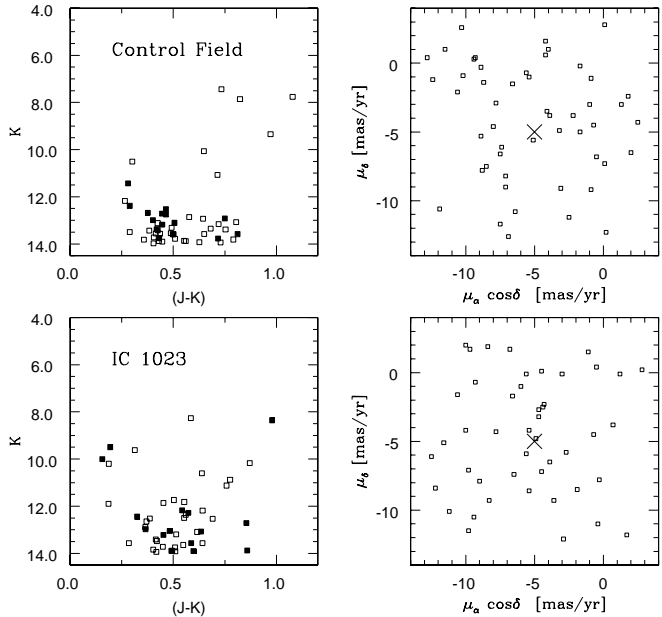}
\caption{Same of Fig.~6, but for IC1023} 
\end{figure}

\subsection{ESO282SC26}
This target is discussed in Fig.~16. We considered a $30^{\prime} \times 30^{\prime}$ 
area, and we found 2008 stars in the target area, and 2095 stars in the off-set field,
roughly the same number of stars.
The mode of the  proper motion distribution turned out to be about
$\mu_{\alpha} \approx 0.0$ mas yr$^{-1}$ and $\mu_{\delta} \approx -5.0$ mas yr$^{-1}$ in the target and 
$\mu_\alpha~=~-5.0$ and $\mu_\delta~=~0.0$ mas yr$^{-1}$ in the field.
Interestingly, in  the VPDs we can see that the target is more populated than the field,
and therefore this is a very promising object to be considered.
At odds with all the previous cases, the CMD is very interesting.
The CMD of the target and the field are at first glance very similar, although
the target possesses a more populated blue sequence than the field. 
The red part of the CMD is somewhat more different, since the cluster
have more red stars within 1 SPM3 $\sigma$, and the distribution
of these stars closely resembles a Red Giant Branch feature.\\
We therefore decided to try an isochrone fit through the stars having proper
motion within 1 SPM3 $\sigma$ from the mode of the distributions
and located within 7.5 arcmin from the object nominal center to limit
the effect of field star contamination.\\
This is shown in Fig.~17, where the isochrone 
is for solar abundance and for an age of 1.3 Gyr, 
and has been taken from Bonatto et al (2004).\\

\noindent
The fit is very good, although some scatter is still present. As a by-product,
we find that the cluster is 1.4 kpc away from the Sun, and it is located inside the solar
ring. Therefore we suggest that ESO282SC26 is a probable open 
cluster which deserves further investigation.

\begin{figure}
\centering
\includegraphics[width=9cm]{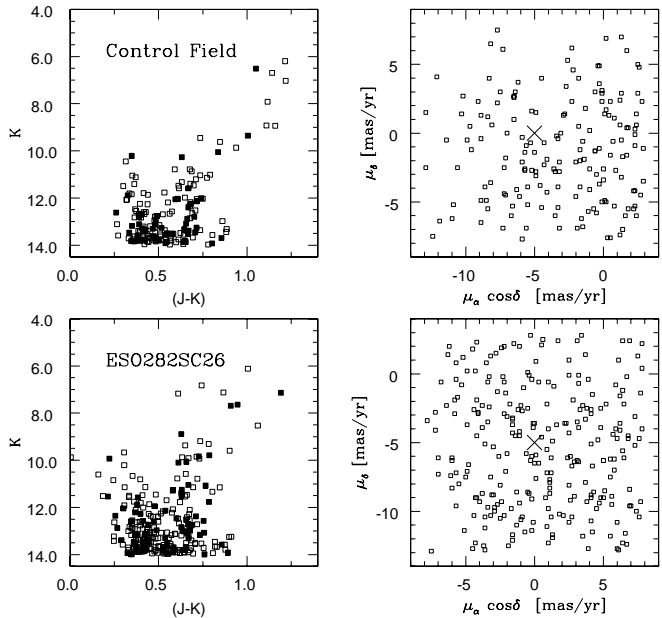}
\caption{Same of Fig.~6, but for ESO282SC26} 
\end{figure}

\begin{figure}
\centering
\includegraphics[width=9cm]{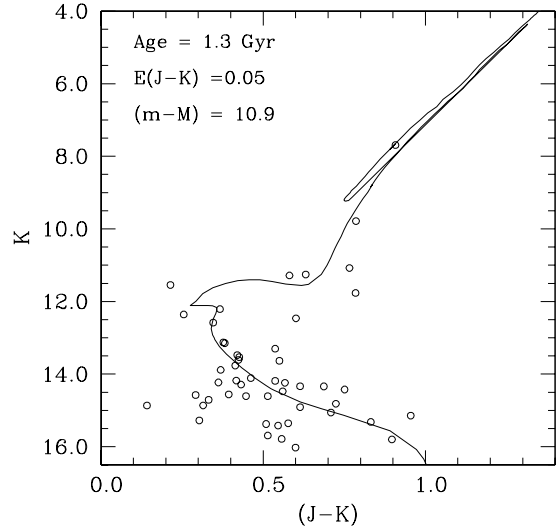}
\caption{2MASS CMD of common proper motion stars in the field of ESO282SC26
within 7.5 arcmin from the cluster nominal center.} 
\end{figure}

\subsection{ESO464SC09}
This target is depicted in Fig.~18. We considered a $15^{\prime} \times 15^{\prime}$
area, and we found 150 stars in the target area, and 148 stars in the off-set field,
i.e. the target and the field are equally populated.
The mode of the  proper motion distribution turned out to be about
$\mu_{\alpha} \approx 0.0$ mas yr$^{-1}$ and $\mu_{\delta} \approx -5.0$ mas yr$^{-1}$ in the target and 
$\mu_\alpha~=~-5.0$ and $\mu_\delta~=~0.0$ mas yr$^{-1}$ in the field, in both cases
with a very large scatter.
This object is very poorly populated, and we suggest we are 
facing a random enhancement of field stars.

\begin{figure}
\centering
\includegraphics[width=9cm]{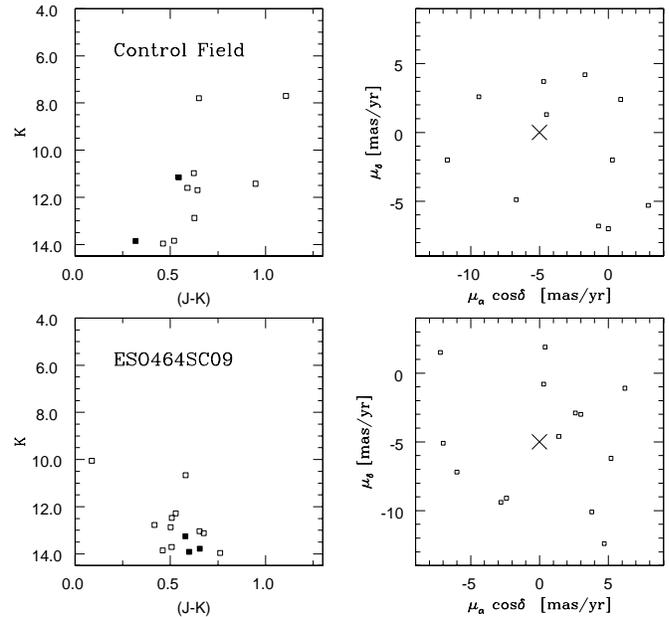}
\caption{Same of Fig.~6, but for ESO464SC09} 
\end{figure}

\section{Conclusions}
In this paper we made use of the SPM3 catalog to assess
the nature of eleven POCRs extracted from Bica et al. (2001) investigation. 
We first demonstrated the capability of the catalog to unravel
a physical group by studying the open cluster Blanco 1.
Then we analyzed VPDs and CMDs of stars selected on a common proper motion
basis.\\

However, within the SPM typical uncertainty,  a close scrutiny of the VPDs and CMDs
clarifies that only one -ESO282SC26- out of eleven candidates looks like
a physical group, and for it we provide estimates 
of its fundamental parameters.

\begin{acknowledgements}
The authors thanks the anonymous referee for very useful suggestions
which contributed to significantly improve the paper presentation.
The work of GC is supported by {\it Fundaci\'on Andes}.
This study made use of Simbad and WEBDA. 
 \end{acknowledgements}

\end{document}